\documentclass[twocolumn,showpacs,preprintnumbers,amsmath,amssymb]{revtex4-1}

\usepackage{graphicx}
\usepackage{amssymb}
\usepackage{hyperref}

\begin{document}

\title{Local character of the highest antiferromagnetic Ce-system CeTi$_{1-x}$Sc$_x$Ge}

\author{J.G. Sereni, P. Pedrazzini, M. G\'omez Berisso, A. Chacoma, S. Encina}

\address{Low Temperature Division, CAB-CNEA and CONICET, 8400 San Carlos de Bariloche, Argentina}

\author{T. Gruner, N. Caroca-Canales, C. Geibel}

\address{Max-Planck Institute for Chemical Physics of Solids, D-01187
Dresden, Germany}

\date{\today}

\begin{abstract}

{The highest antiferromagnetic (AFM) temperature in Ce based
compounds has been reported for CeScGe with $T_N = 47$\,K, but its
local or itinerant nature was not deeply investigated yet. In
order to shed more light into this unusually high ordering
temperature we have investigated structural, magnetic, transport
and thermal properties of CeTi$_{1-x}$Sc$_x$Ge alloys within the
range of stability of the CeScSi-type structure: $0.25 \leq x \leq
1$. Along this concentration range, this strongly anisotropic
system presents a complex magnetic phase diagram with a continuous
modification of its magnetic behavior, from ferromagnetism (FM)
for $0.25 \leq x \leq 0.50$ (with 7\,K$ \leq T_C \leq 16$\,K) to
AFM for $0.60 \leq x \leq 1$ (with 19\,K$\leq T_N \leq 47$\,K).
The onset of the AFM phase is associated to a metamagnetic
transition with a critical field increasing from $H_{cr} = 0$ at
$x\approx 0.55$ to $\approx 6$\,Tesla at $x = 1$, coincident with
an increasing contribution of the first excited crystal electric
field doublet. At a critical point $x_{cr} = 0.65$ a second
transition appears at $T_L \leq T_N$. In contrast to observations
in itinerant systems like CeRh$_2$Si$_2$ or CeRh$_3$B$_2$, no
evidences for significant hybridization of the 4f electrons at
large Sc contents were found. Therefore, the exceptionally large
$T_N$ of CeScGe can be attributed to an increasing RKKY
inter-layer interaction as Sc content grows.}

\end{abstract}

%\pacs{75.20.Hr, 71.27.+a, 75.30.Kz, 75.10.-b\\
%$^*$ E-mail-address: jsereni@cab.cnea.gov.ar}

\maketitle

\section{Introduction}

Magnetic order in Cerium compounds can be found along four decades
of  temperature, from 1.6\,mK in CMN \cite{CMN} up to 115\,K in
CeRh$_3$B$_2$ \cite{Malik}, though most transitions occur between
a few degrees and $T_{ord} \approx 12$\,K. Different magnetic
behaviors characterize different temperature ranges of magnetic
order. While magnetic interactions between localized Ce-$4f$
moments are observed for $T_{ord} < 12$\,K \cite{Bauer91}, in the
cases where $T_{ord}$ exceeds that temperature, two types of
behaviors can be distinguished. The one with localized Ce-$4f$
moments is related to the rare cases of Ce binary compounds formed
in BCC structure. This is the crystalline structure with the
highest symmetry that allows to have the quartet $\Gamma_8$ as
ground state. However, its four fold degeneracy is reduced by
undergoing different types of transitions mostly connected with
structural modifications. Those compounds are: CeZn ($T_N=30$\,K);
CeTl ($T_N=25.5$\,K); CeMg ($T_C=20$\,K); CeCd ($T_C=16.5$\,K) and
CeAg ($T_N=15.6$\,K) \cite{Hand91}.

The ternary Ce-based compounds belonging to the other group
present evidences of itinerant character \cite{DeLong91}. Among
them, the outstanding case is the mentioned CeRh$_3$B$_2$ that
shows the highest ordering temperature with $T_{ord} = 115$\,K
\cite{Malik}. The following highest ordering temperature has been
reported for CeScGe with $T_{ord} = 47$\,K \cite{Canfield91}, but
the local or itinerant nature of its magnetic state is under
discussion. Within this group, one of the most studied compounds
is CeRh$_2$Si$_2$, which shows a $T_{ord}=36$\,K
\cite{Godart83,CeRh2Si2}. Also in this case evidences for local
and itinerant magnetic character were equally claimed by different
authors \cite{8kawa,5settai97,Beri}.

Such ambiguity is not unusual in Ce compounds and it has been
discussed since the early studies on pure Ce under pressure and
identified as the local-itinerant dilemma of Ce-$4f$ electrons
\cite{Mackin}, not being completely elucidated yet. The eventual
itinerant character of the $4f$ orbitals was compared to the
U-$5f$ behavior \cite{DeLong91}, whose compounds frequently show
high $T_{ord}$ values with clear itinerant characteristics arising
from the extended character of the $5f$ orbitals. In Ce-$4f$
systems, however, itineracy is related to some degree of
hybridization with the conduction band and the consequent
weakening of the $4f$ effective magnetic moment $\mu_{eff}$. The
question arises whether there is an optimal combination of weak
$4f$-band hybridization that enhances the inter-site RKKY
interaction by increasing the in-site $4f$-band exchange without
reducing $\mu_{eff}$ significantly.

The most convenient systems to investigate this problem are those
showing high $T_{ord}$ accompanied by experimental evidences where
itineracy or localization character may emerge. In the case of
CeScGe, it was formerly reported as a ferromagnet (FM)
\cite{Canfield91} that usually implies local $4f$ character.
However, band-structure calculations \cite{Shigeoka} suggested
that FM and antiferromagnetic (AFM) states compete in energy based
in an itinerant character. Later on, a more detailed study on
CeScSi and CeScGe \cite{Singh} recognized both compounds as AFM.
Thus, CeScGe presents the best characteristics for the proposed
investigation.

A further aspect has to be taken into account for a proper
knowledge of the ordered phase at those temperatures, which is the
role of the excited crystal-electric-field (CEF) levels. This
contribution can be properly evaluated by tracing $T_{ord}$ from
relatively low values, where only a doublet with local-$4f$
character contributes to the magnetic ground state. These
conditions are fulfilled by CeTi$_{1-x}$Sc$_x$Ge alloys, that
cover an extended range of transition temperatures from $T_{ord}
\approx 7$\,K in CeTi$_{0.75}$Sc$_{0.25}$Ge to $T_{ord}\approx
47$\,K at the stoichiometric limit CeScGe \cite{Canfield91} . The
lower $T_{ord}$ value is determined by the limit of stability of
the CeScSi-type structure (at $x=0.25$) on the Ti rich side. Below
$x \approx 0.15$ this alloy stabilizes in the CeFeSi-type
structure.

\section{Experimental details and results}

Polycrystalline samples of CeTi$_{1-x}$Sc$_x$Ge with $0.25 \leq x
\leq 1$ were synthesized by arc melting under Ar atmosphere the
nominal amounts of the constituents, purity above 99.99\%. The
samples were turned over and remelted several times to ensure
homogeneity. Then, the samples were placed in a tungsten boat
wrapped with zirconium foil and annealed at 1200\,$^o$C for one
week. The quality of the samples was verified by means of X-ray
powder-diffraction measurements using Cu-K$\alpha_1$ radiation
($\lambda =1.54056\, \rm{\AA}$) in a Stoe-Stadip-MP
diffractometer. The pattern was indexed on the basis of the
tetragonal CeScSi-type structure.

\begin{figure}
\begin{center}
\includegraphics[angle=0, width=0.45 \textwidth, height=1.1\linewidth] {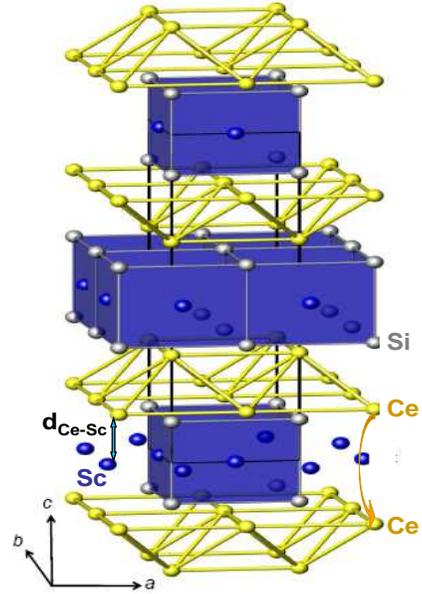}
\end{center}
\caption{(Color online) CeScSi-type structure, showing double Ce
layers (yellow open network) and ligands layers (blue full
network). Left side: $d_{Ce-Sc}$ indicates Ce-Sc spacing. Right
side: curved arrow represents Sc mediated Ce-inter-layers
interaction.} \label{F1}
\end{figure}

Specific heat was measured between 0.5 and 50K using a standard
heat pulse technique in a semi-adiabatic He$^3$ calorimeter. The
magnetic contribution $C_m$ is obtained by subtracting the phonon
contribution extracted from LaTi$_{0.5}$Sc$_{0.5}$Ge.
Magnetization measurements were carried out using a Quantum Design
MPMS magnetometer operating between 2 and 300K, and as a function
of field up to 50kOe. Electrical resistivity was measured between
2\,K and room temperature using a standard four probe technique
with an LR700 resistive bridge.

\subsection{Structural properties}

\begin{figure}
\begin{center}
\includegraphics[angle=0, width=0.5 \textwidth, height=0.75\linewidth] {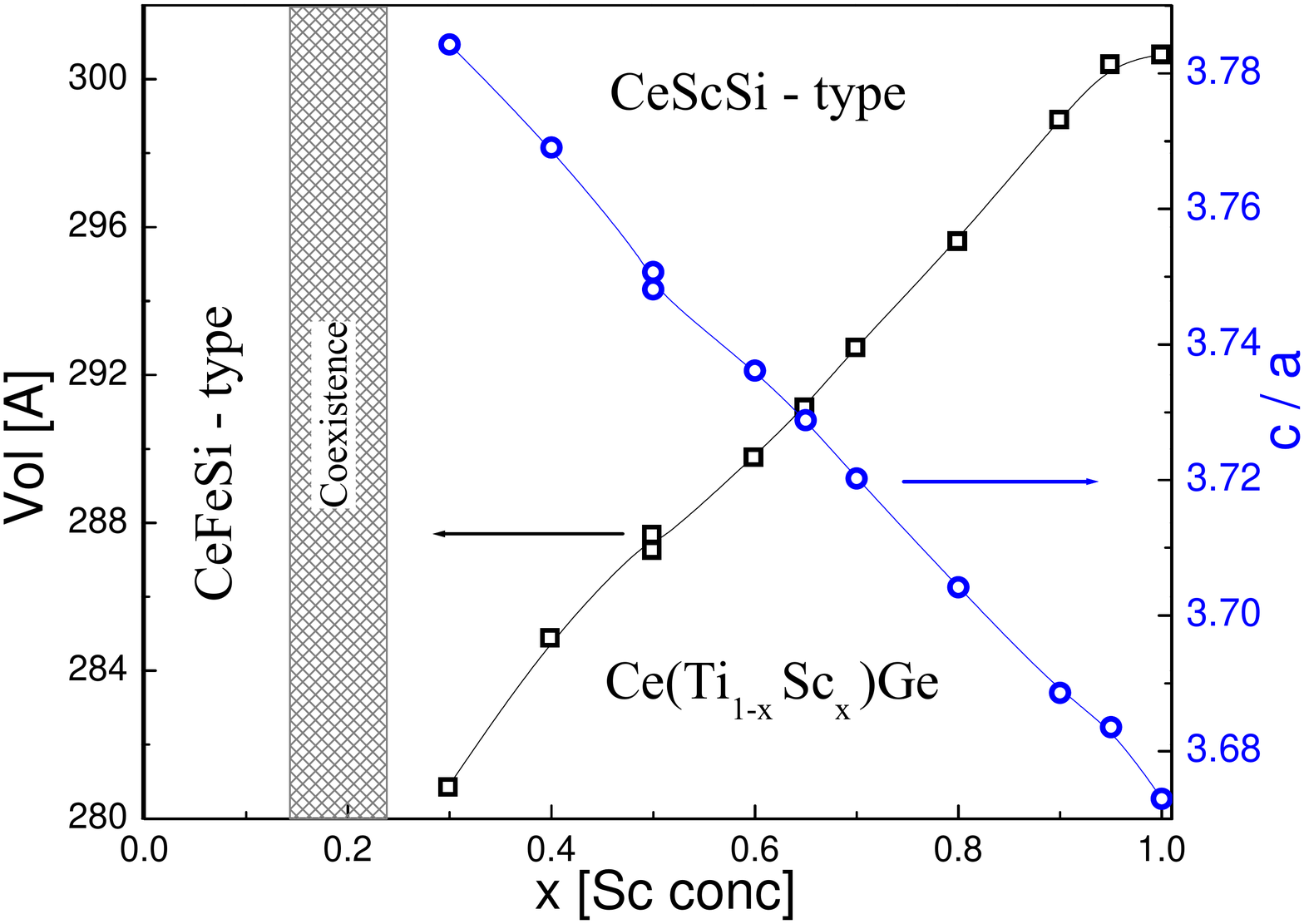}
\end{center}
\caption{(Color online) Unit cell volume variation in
CeTi$_{1-x}$Sc$_x$Ge as a function of Sc content (left axis) and
'c/a' lattice parameters ratio (right axis). The shaded area
indicates the coexistence region of both structures.} \label{F2}
\end{figure}

The CeTi$_{1-x}$Sc$_x$Ge system forms in two related crystal
structure: CeFeSi-type for low Sc content (up to $x=0.15$) and
CeScSi-type beyond $x=0.23$  \cite{Gruner}. In the latter, each
second Ce-double layer is shifted by (1/2, 1/2) with a
rearrangement of Sc and Si atoms, becoming a body centered
tetragonal instead of primitive tetragonal of CeFeSi, see
Fig.~\ref{F1}, with the consequent doubling of the 'c' lattice
parameter.

The unit cell volume dependence on Sc concentration and the 'c/a'
ratio of the tetragonal lattice parameters are shown in
Fig.~\ref{F2}. The increase of the unit cell volume can be
explained by the larger atomic volume of Sc with respect to Ti,
whereas the reduction of the 'c/a' ratio is due to the increase of
the 'a' parameter because 'c' remains practically unchanged. This
indicates that an expansion in the basal plain of the tetragonal
structure occurs whereas the inter-layer distances in the 'c' axis
direction are slightly affected.

\subsection{Magnetic susceptibility}

\begin{figure}
\begin{center}
\includegraphics[angle=0, width=0.5 \textwidth, height=0.75\linewidth] {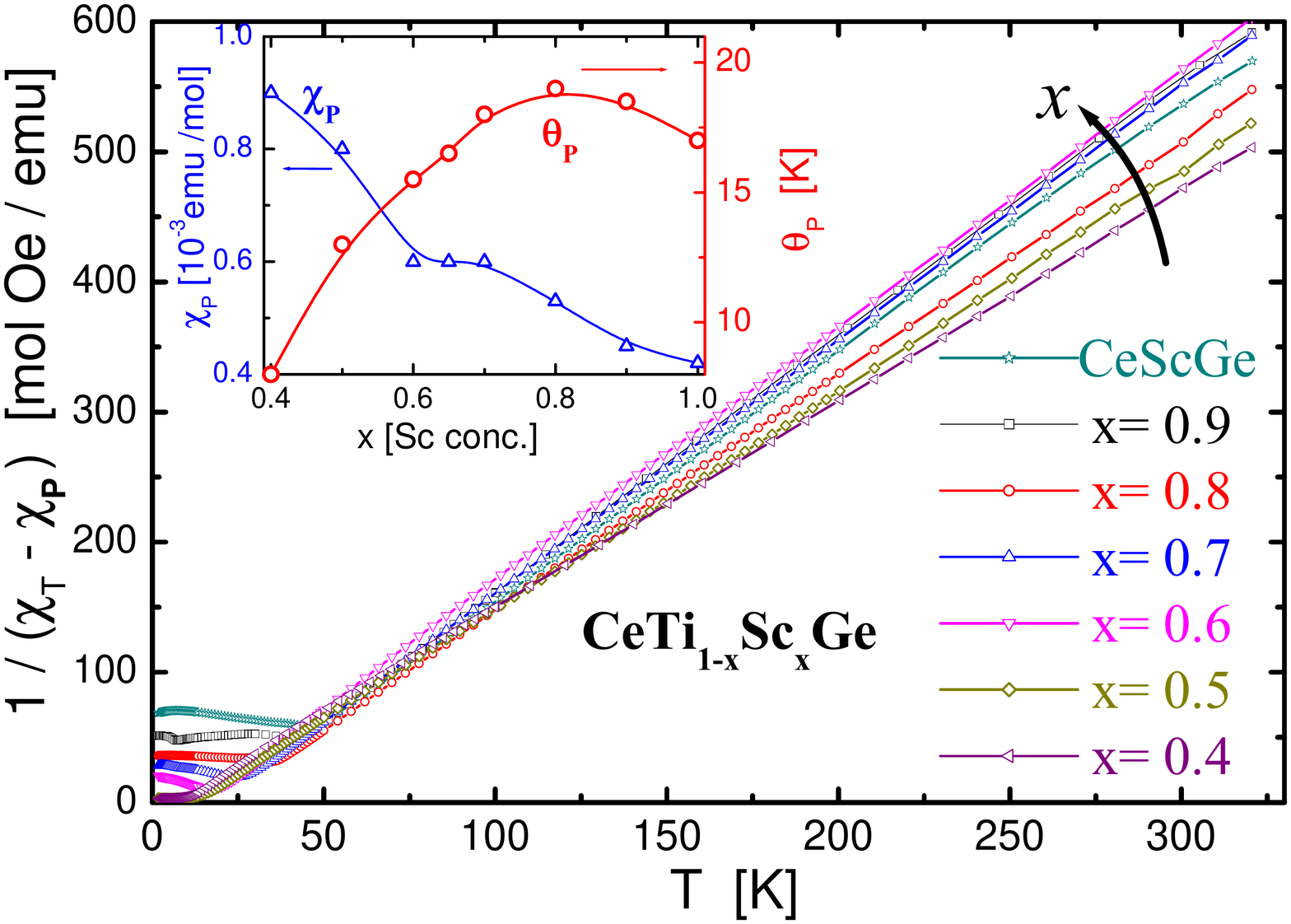}
\end{center}
\caption{(Color online) Inverse high temperature magnetic
susceptibility measured in a field of $H=10 kOe$, after
subtracting a Pauli $\chi_P$ contribution. Inset: $\chi_P(x)$
contribution (left axis) and paramagnetic temperature
$\theta_P(x)$ (right axis). } \label{F3}
\end{figure}

In Fig.~\ref{F3}, the inverse of the high temperature magnetic
susceptibility ($1/\chi$) is presented after subtracting a Pauli
type paramagnetic contribution $\chi_P$. This temperature
independent contribution, already reported for CeScGe
\cite{Uwatoko}, is observed to decrease from $\chi_P = 0.9
10^{-3}$ at $x = 0.4$ to $0.47 10^{-3}$emu/Oe\,mol at $x = 1$, as
shown in the inset of Fig.~\ref{F3}. The $1/\chi$ dependence with
concentration (evaluated for $T>50$\,K) indicates a decrease of Ce
effective magnetic moment from $\mu_{eff}(x) \approx 2.25\mu_B$ at
$x = 0.4$ to $\approx 2\mu_B$ at $x = 0.7$. Beyond that
concentration $\mu_{eff}(x)$ remains practically unchanged.
Notably, the paramagnetic temperature $\theta_P (x)$ is positive
along the full concentration range and even increases from
$\theta_P \approx 8$\,K at $x = 0.4$, up to 19\,K at $x = 0.8$. At
that concentration it slightly decreases down to 17\,K at $x=1$ as
shown in the inset of Fig.~\ref{F3}.

\begin{figure}
\begin{center}
\includegraphics[angle=0, width=0.5 \textwidth, height=0.7\linewidth] {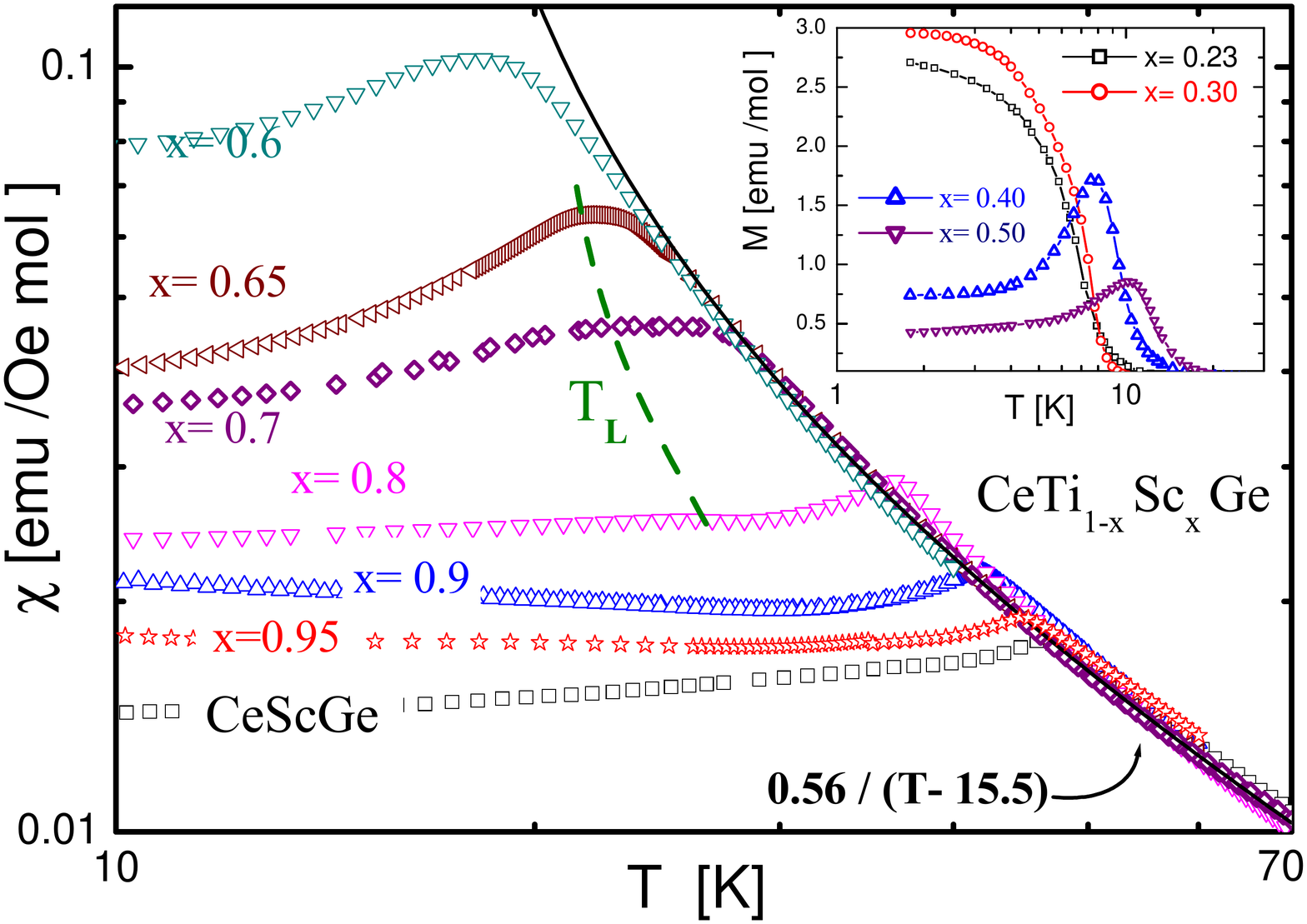}
\end{center}
\caption{(Color online) Low field ($H = 100 Oe$) magnetic
susceptibility in a double logarithmic representation. Full curve
is a reference for the paramagnetic phase (see the text) and
dashed curve indicates the position of the second transition at $T
= T_L$. Inset $M(T)$ measurements for $x \leq 0.50$ samples.}
\label{F4}
\end{figure}

A detailed variation of $\chi(T)$ around $T_{ord}$ is shown in
Fig.~\ref{F4} in a double logarithmic representation within the
range where a maximum of $\chi(20\leq T_{ord} \leq 50\,K)$
indicates the AFM character of the transition for samples with
$x\geq 0.60$. The samples with $x\leq 0.5$ are included in the
inset of Fig.~\ref{F4} using a more extended $M(T)$ scale because
of a FM contribution. The typical $M(T)$ dependence of a FM is
observed for $x=0.25$ and 0.3 alloys, whereas those with $x\geq
0.4$ show an incipient AFM component even in a field cooling
procedure. The thermal dependence on the paramagnetic phase is
compared in the main figure with a general Curie-Weiss function
$\chi(T) = 0.56 /(T-15.5)$emu/Oe\,mol (notice the positive
$\theta_P=15.5$\,K) to remark the stability of the paramagnetic
phase along the full concentration range. On the contrary, the
ordered phase shows a continuous variation of $\chi(T<T_N)$
including the presence of a weak maximum at $T=T_L$ between $0.65
\leq x \leq 0.80$.

Those different behaviors can be sorted into three ranges of Sc
concentration: i) for $x\leq 0.6$ the $M(T)$ dependence indicate
an increasing mixture of FM and AFM components in the GS, with the
ordering temperature increasing from $T_{ord}= 7$K at $x = 0.25$
up to 16K at $x = 0.60$. ii) At $x=0.65$ the $\chi(T)$ maximum
displays two shoulders which can be distinguished after a detailed
analysis of the $\chi(T)$ curvature, i.e. its second derivative
$\partial^2\chi / \partial T^2$. This feature reveals that the $x
= 0.65$ concentration is placed very close to a critical point.
iii) Between $0.65 \leq x \leq 0.8$, a slight kink appears at $T =
T_L (x)$, below $T_N$, as seen in Fig.~\ref{F4}. Within this range
of concentration, $T_N$ increases from 19\,K at $x = 0.65$ up to
35\,K at $x = 0.8$, whereas $T_L(x)$ increases from 18\,K up to
26\,K. iv) Above $x = 0.9$, $T_N$ keeps growing from 38\,K at $x =
0.9$ up to 47\,K at $x = 1$, whereas $T_L(x)$ is hardly seen in
$\chi(T)$ measurements.

\subsection{Electrical resistivity}

\begin{figure}
\begin{center}
\includegraphics[angle=0, width=0.5 \textwidth, height=0.7\linewidth] {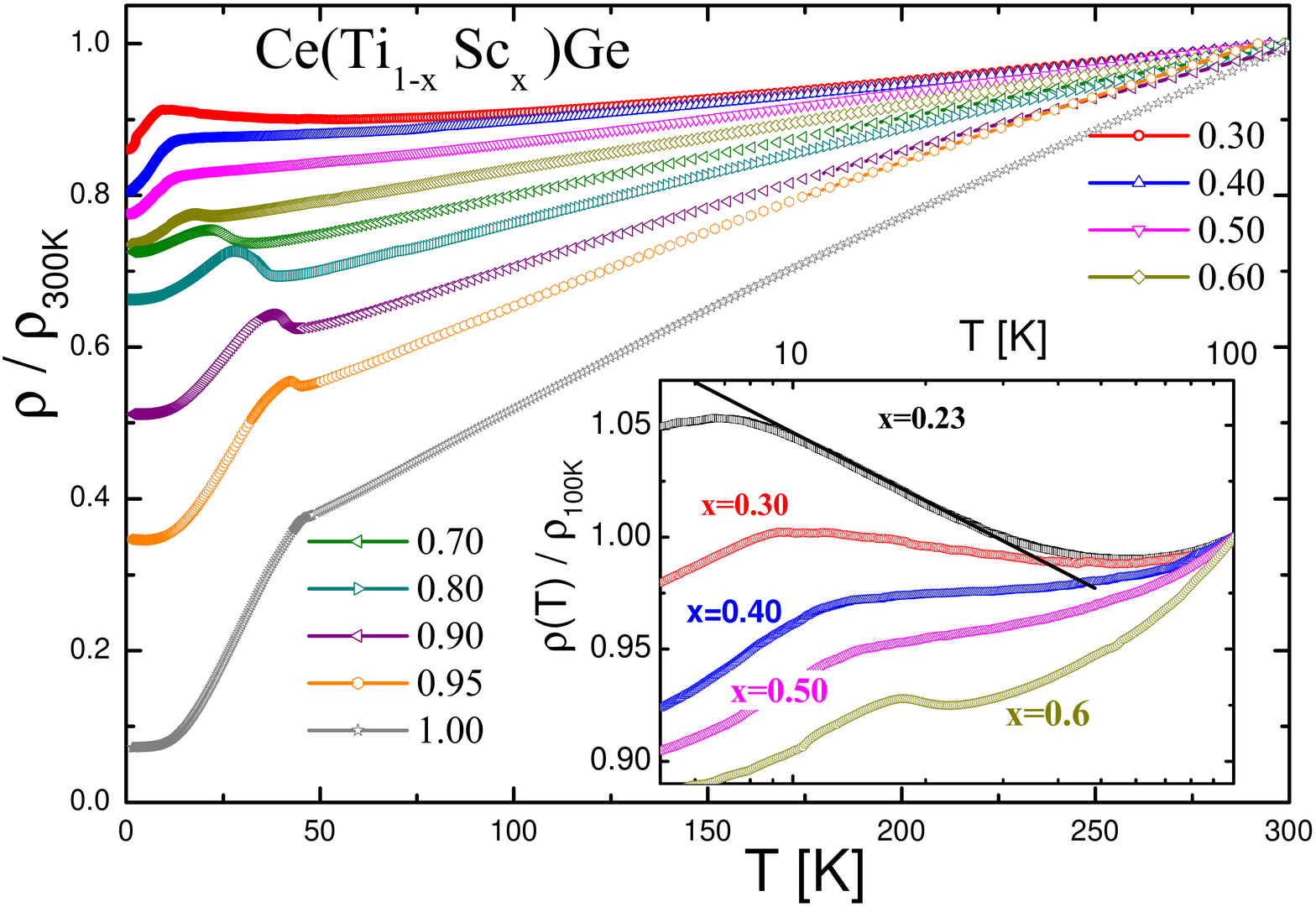}
\end{center}
\caption{(Color online) Electrical resistivity normalized at
300\,K of all measured samples. Inset: detail of the logarithmic
$T$ dependence of samples between $0.23 \leq x \leq 0.4$
normalized at 100\,K for better comparison.} \label{F5}
\end{figure}

The thermal dependence of the electrical resistivity $\rho$,
normalized at 300\,K, are presented in Fig.~\ref{F5} for samples
between $0.30 \leq x \leq 1.0$. The continuous increase of the
$\rho(T)$ slopes indicates the weakening of Kondo type scattering
with growing Sc content. In the inset of that figure, a detail of
the logarithmic dependence of samples on the Ti-rich side ($0.23
\leq x \leq 0.60$) normalized at 100\,K are collected to better
analyze the role of the Kondo effect. A logarithmic increase of
$\rho(T)$ approaching the magnetic transition from high
temperature is observed at $x=0.23$ as an indication of Kondo
scattering contribution. However, the intensity of this electronic
scattering rapidly decreases with Sc concentration till to vanish
at $x=0.60$. At that concentration, the characteristic resistivity
upturn for an AFM transition emerges. This anomaly develops up
$x=0.80$ and then decreases again as the system reaches its
stoichiometric CeScGe limit.

The temperature of $\rho(T_{ord})$ maxima are in agreement with
the transitions observed from magnetic measurements and its shape
changes according to the modification of the magnetic nature of
the GS from FM to AFM. However the width of the $\rho{T_ord}$
anomaly involves the temperature range between $T_N$ and $T_L$.
The weakening of the $\rho(T)$ anomaly approaching the
stoichiometric limit can be attributed to the variation of the
Fermi level within the energy gap associated to the AFM state. As
Ti$^{4+}$ atoms are replaced by Sc$^{3+}$ ones, the chemical
potential decreases and eventually approaches the lower energy
side of the gap. This simple picture has to be considered within
the complexity of the Fermi Surface in a real system with many
conducting electrons.

\subsection{Specific heat}

\begin{figure}
\begin{center}
\includegraphics[angle=0, width=0.5 \textwidth, height=0.8\linewidth] {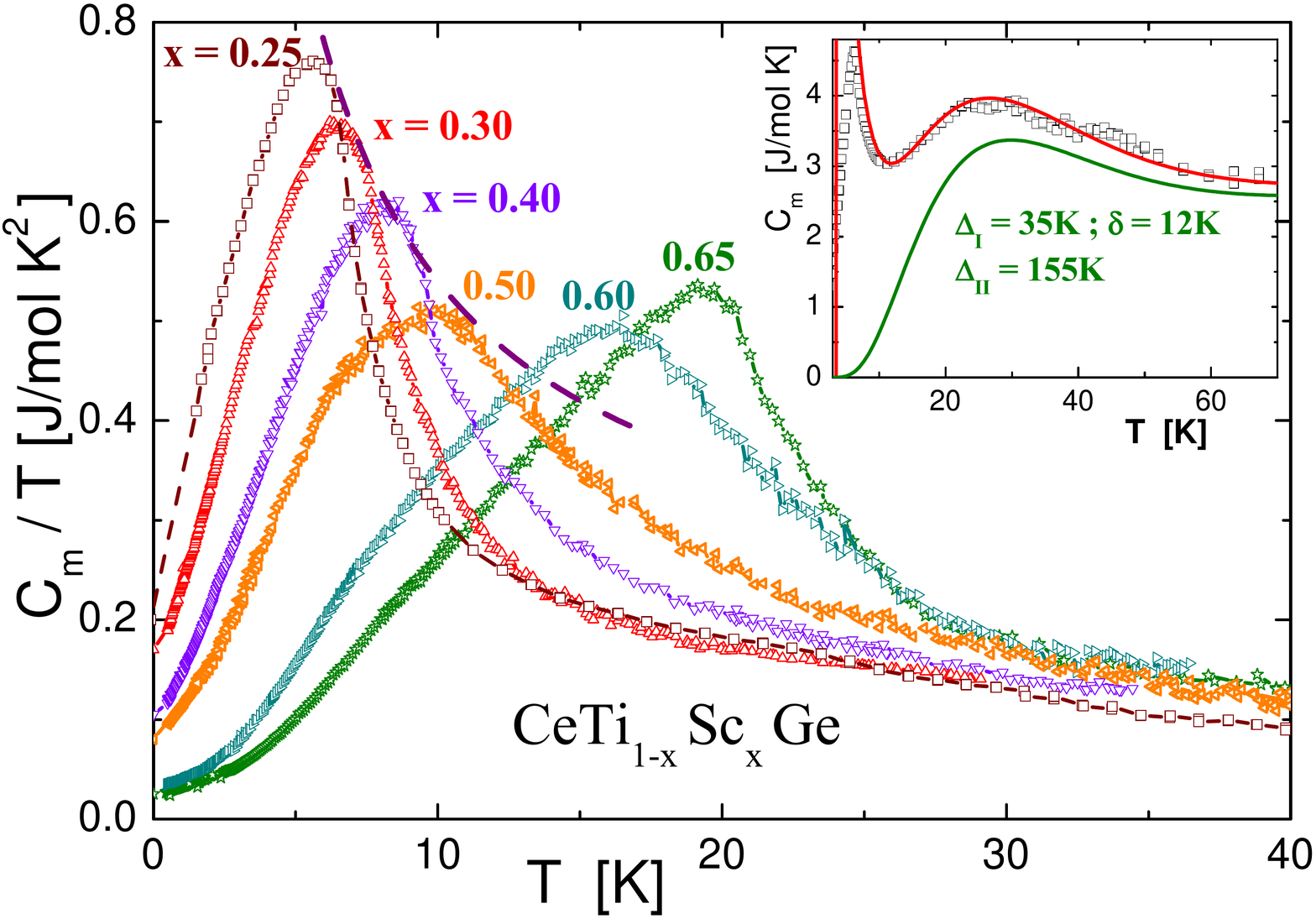}
\end{center}
\caption{(Color online) Magnetic contribution to the specific heat
divided temperature within the $0.25\leq x \leq 0.65$ range. Doted
curve indicates the $1/T_{ord}$ decreasing trend of the maximum
below x = 0.5. Inset: fit of $C_m(T)$ for sample $x=0.25$ at high
temperature to evaluate the CEF spectrum, see the text.}
\label{F6}
\end{figure}

Specific heat results show several peculiar features along the
full concentration range in coincidence with the different regimes
observed in magnetic measurements. The results obtained on the
$0.25 \leq x \leq 0.65$ alloys are presented in Fig.~\ref{F6} as
$C_m/T$, where $C_m$ indicates the magnetic contribution to the
specific heat after phonon subtraction. Within the experimental
dispersion, the maxima of $C_m(x)/T$ coincide with those observed
in $M(T)$. A common feature along this concentration range is the
lack of a $C_m$ jump at $T=T_{ord}$. Instead, a broadened mean
field transition is observed, followed by a tail in
$C_m/T(T>T_{ord})$ that reveals a significant contribution of
magnetic correlations as precursors of the magnetic transition.
This tail becomes more extended in temperature with increasing
$T_{ord}(x)$ up to $x=0.50$. Samples with $x=0.60$ and $x=0.65$
recover the slope above $T_{ord}$ together with a moderate
increase of the $C_m/T$ maximum. This change of tendency coincides
with the modification of the magnetic structure around $x=0.50$.
In this concentration range, the maximum of $C_m/T$ decreases as
$\propto 1/T_{ord}$. Although such a decrease is a consequence of
plotting $C_m(T)/T$, it reveals that $C_m(T_{ord})$ slightly
decreases with decreasing $T_{ord}$ and does not extrapolate to
zero according to the law of corresponding states.

\begin{figure}
\begin{center}
\includegraphics[angle=0, width=0.5 \textwidth, height=1.4\linewidth] {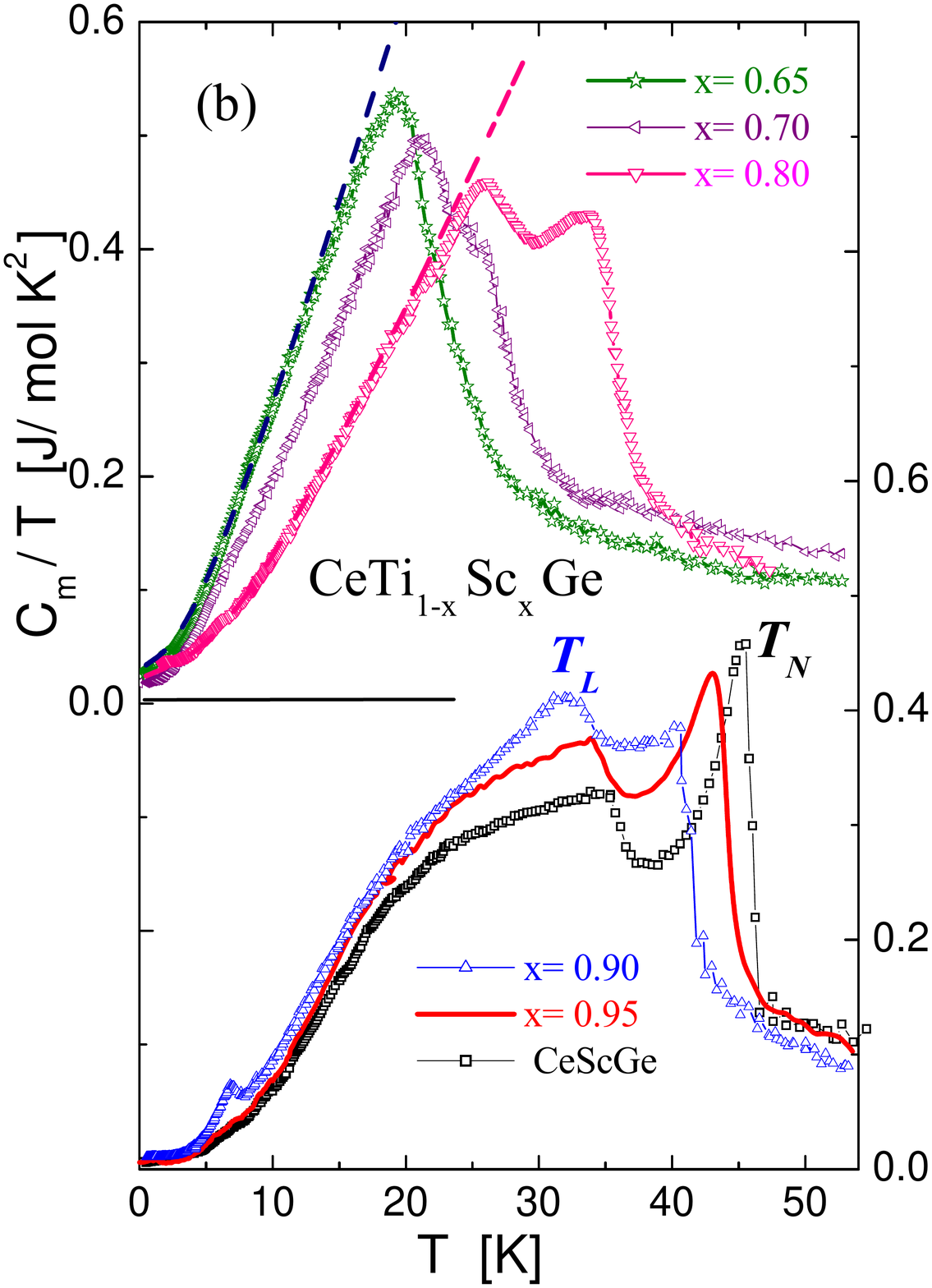}
\end{center}
\caption{(Color online) Magnetic contribution to the specific heat
in the $0.65 \leq x \leq 0.80$ range (left axis), and the $0.90
\leq x \leq 1$ samples shifted for clarity (right axis). Dashed
curves represent the fits on the ordered phase of samples $x=0.65$
and 0.80 (see the text).} \label{F7}
\end{figure}

An increase of the low temperature curvature in $C_m/T(T)$
indicates a progressive opening of a gap of anisotropy ($\Gamma$)
in the magnon spectrum with concentration. To evaluate that
evolution, the low temperature $C_m/T$ dependence was analyzed
using the function  $C_m/T= \gamma_0 + B*T*exp(-\Gamma /T)$ as
shown in Fig.~\ref{F7} for samples with $x=0.65$ and 0.8. The
values computed for those concentrations are $\Gamma=7$\,K and
$\Gamma = 15$\,K respectively. Concerning the $\gamma_0(x)$
dependence, it drops from $\gamma_0 \approx 0.22$\,J
mol$^{-1}$K$^{-2}$ at $x=0.25$ down to $\approx 0.02$\,J
mol$^{-1}$K$^{-2}$ at $x=0.60$.

The main characteristic of the alloys with $x > 0.65$ is the split
of $C_m/T$ into two maxima, in agreement with the $\chi(T)$
results. While the lower transition (at $T=T_L$) shows a cusp-like
anomaly between $x=0.7$ and 0.9, the upper one ($T_N$) is
associated to a jump in $C_m/T(T_N)$. The cusp at $T=T_L$ can be
related to the vicinity of the critical point accounting that in
these policrystalline samples any eventual first order transition
could be broadened in temperature because of the random direction
distribution of crystals and some eventual concentration
inhomogeneities.

For $x\geq 0.9$, a further change in $C_m/T(T)$ is observed as
depicted in the lower part of Fig.~\ref{F7}. The broad shoulder of
$C_m/T(T)$ in samples $0.9\leq x \leq 1$ can be attributed to the
increase of the GS degeneracy from $N_{eff} =2$ to 4 as the
contribution of the first exited CEF doublet gradually increases.
Also both transitions change their shapes because, while the one
at $T_L$ transforms into a step like anomaly, that at $T_N$
becomes sharper and grows significantly. Notice that the $\Delta
C_m(T_N)$ jump in $x=1$ is $\approx 16$\,J mol$^{-1}$K$^{-1}$, in
between the values predicted in a mean field approximation for a
doublet ($\Delta C_m =1.5 R = 12.5$\,J mol$^{-1}$K$^{-1}$) and a
quartet ($\Delta C_m =2.2 R \approx 18$\,J mol$^{-1}$K$^{-1}$)
\cite{Morphol}. The small anomaly observed in $x=0.9$ at $T\approx
7$K can be attributed to an extrinsic contribution of a small
amount of Ce-oxide.

\section{Discussion}

\begin{figure}
\begin{center}
\includegraphics[angle=0, width=0.5 \textwidth, height=1.4\linewidth] {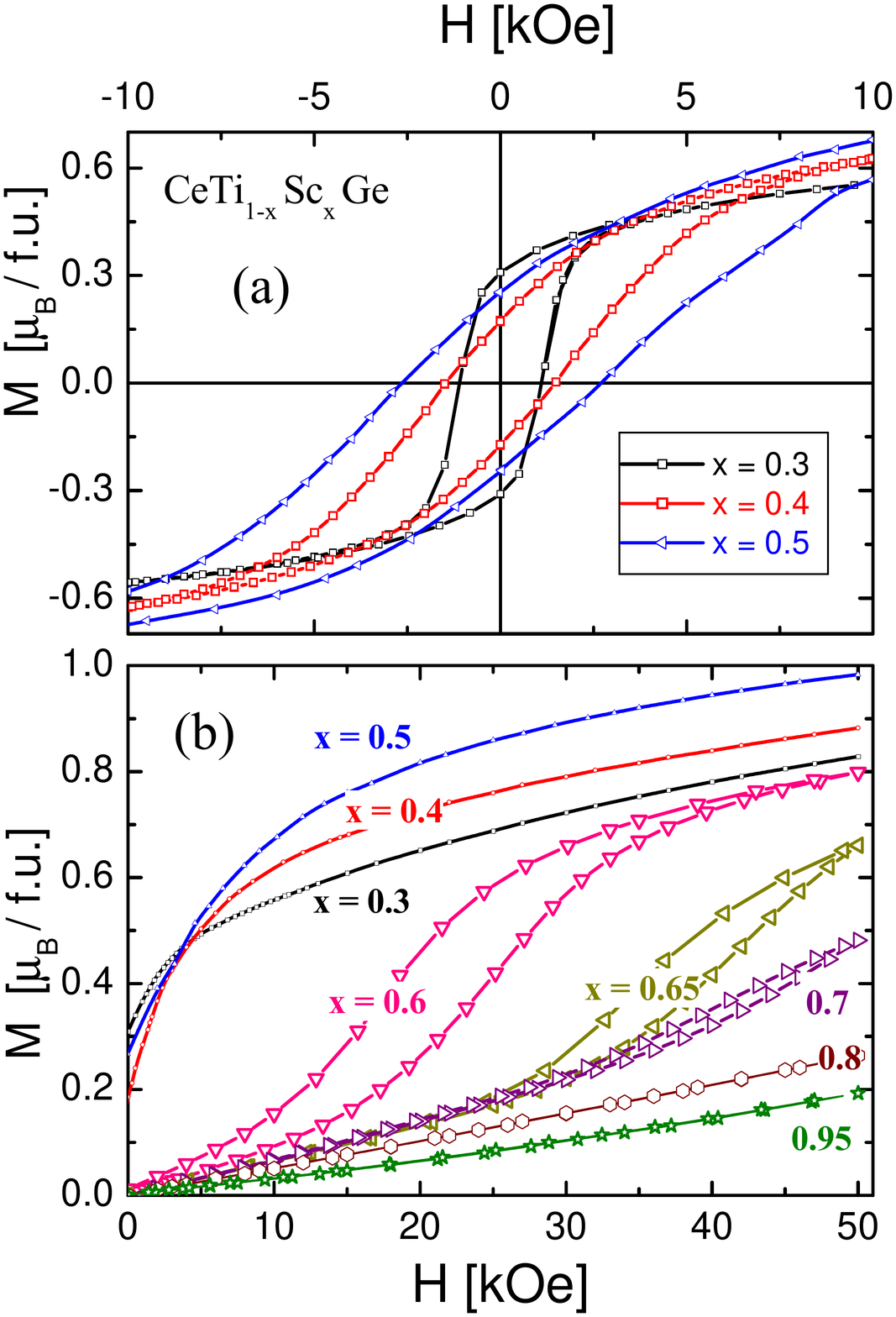}
\end{center}
\caption{(Color online) Magnetization measurements at $T =
1.8$\,K: (a) hysteresis loops centered at $H = 0$ for the FM $0.3
\leq x \leq 0.5$ alloys. (b) Magnetization curves up to $H =
50$kOe, showing a metamagnetic transformation in samples with $x
\geq 0.6$.} \label{F8}
\end{figure}

\subsection{Magnetization}

The $M(H)$ hysteresis loops measured at $T=1.8$\,K on the alloys
with $x\leq 0.5$ reveal the FM character of the ordered phase (see
Fig.~\ref{F8}a) concomitant with the increasing value of $M(x)$
measured at $H=50$\,kOe presented in Fig.~\ref{F8}b. The
saturation magnetization ($M_{sat}$), extracted from
Fig.~\ref{F8}b as $M(x,H)= M_{sat}\times (1-a/H)$, increases from
1.04\,$\mu_B/f.u.$ for $x=0.30$ up to 1.15\,$\mu_B/f.u.$ for
$x=0.50$. At $x=0.6$ $M_{sat}$ decreases to 1\,$\mu_B/f.u.$ for
$x=0.60$, though this value is extracted from a reduced fitting
range. Similarly, the coercive field increases from $1.1$\,kOe for
$x=0.30$ up to 2.6kOe for $x=0.50$.

At $x=0.60$, the spontaneous FM magnetization is replaced by a
small susceptibility, indicating a transition to an AFM state.
Coincidentally, a metamagnetic transition (MMT) appears with the
critical field $H_{cr}$ increasing with Sc concentration up to our
experimental limit of $H=50$kOe with an initial ratio of $\partial
H_{cr}/\partial x = 2.2$\,kOe/Sc$\%$. $M(H)$ measurements
performed on stoichiometric CeScGe \cite{Uwatoko,Singh} report the
MMT transition around $H_{cr} \approx 60$\,kOe with a weak
associated hysteresis. Notably, the area of the hysteresis loop
decreases as $H_{cr}(x)$ increases, see Fig.~\ref{F8}b, becoming
almost irrelevant at the CeScGe limit \cite{Singh}.

\subsection{Evaluation of the Crystal Electrical Field splitting}

In order to have a qualitative evaluation of the energy of the
excited CEF levels we have analyzed the $C_m(T)$ dependence of the
$x=0.25$ sample which shows the lowest ordering temperature. For
that purpose we have fit the experimental data taking into account
that the CEF splits the six fold Hund's rule multiplet, originated
in the $J=5/2$ orbital momentum of Ce, into three Kramer's
doublets. Due to the eventual hybridization (i.e. Kondo effect)
acting on the first excited level, the standard Schottky anomaly
($C_{Sch}$) may not describe the $C_m(T)$ dependence properly.
Since a proper fit including usual hybridization effects
($V_{cf}$) between conduction and $4f$ states acting on each
excited level requires complex calculation protocols
\cite{Aligia13}, a simple approach to mimic a broadening of the
levels was applied. For this procedure the first CEF exited
doublet at $\Delta_{\rm I}$ is described using single Dirac levels
equally distributed in energy around the nominal value of a non
hybridized level. This method requires the strength of the
hybridization to be smaller than $\Delta_{\rm I}$, i.e. $V_{cf} <
\Delta_I/k_B$. To properly account for the total magnetic
contribution to the entropy, the second excited doublet (at
$\Delta_{\rm II}$) was included, without accounting for eventual
hybridization effects since $\Delta_{II}$ largely exceeds the
temperature range of this analysis. The applied formula is:
\begin{equation}
C_{Sch}(T)= \Sigma_i A_i*[(\Delta_i/T)/\cosh(\Delta_i/T)]^2
\end{equation}
The GS contribution was included after fitting the tail of $C_m$
at $T>T_{ord}$ with an arbitrary $f(a/T)$ function, being the
total contribution to the specific heat: $C_{tot} = f(a/T) +
C_{Sch}(T)$. In the inset of Fig.~\ref{F6} the result of this fit
to the experimental data is shown, including the detail of the
$C_{Sch}$ function. The computed parameters are $\Delta_I=35$\,K
and $\Delta_{\rm II}=155$\,K, with an effective broadening of the
first CEF doublet evaluated as $\delta = 12$\,K. These results
were checked with the evaluation of the magnetic entropy involved
in this Schottky anomaly, which for two excited doublets reaches
$R\ln3$. This analysis cannot be applied for higher Sc
concentrations because the increase of $T_{ord}$ progressively
approaches $\Delta_{\rm I}$. Although no significant changes are
expected in the CEF levels splitting, the reduction of the
electrical charges at the transition metal sites from Ti$^{4+}$
[Ar4s$^2$3d$^2$] to Sc$^{3+}$ [Ar4s$^2$3d$^1$] may affect the
strength of the CEF with a consequent variation of $\Delta_{\rm
I}$.

\subsection{Entropy}

\begin{figure}
\begin{center}
\includegraphics[angle=0, width=0.5 \textwidth, height=0.8\linewidth] {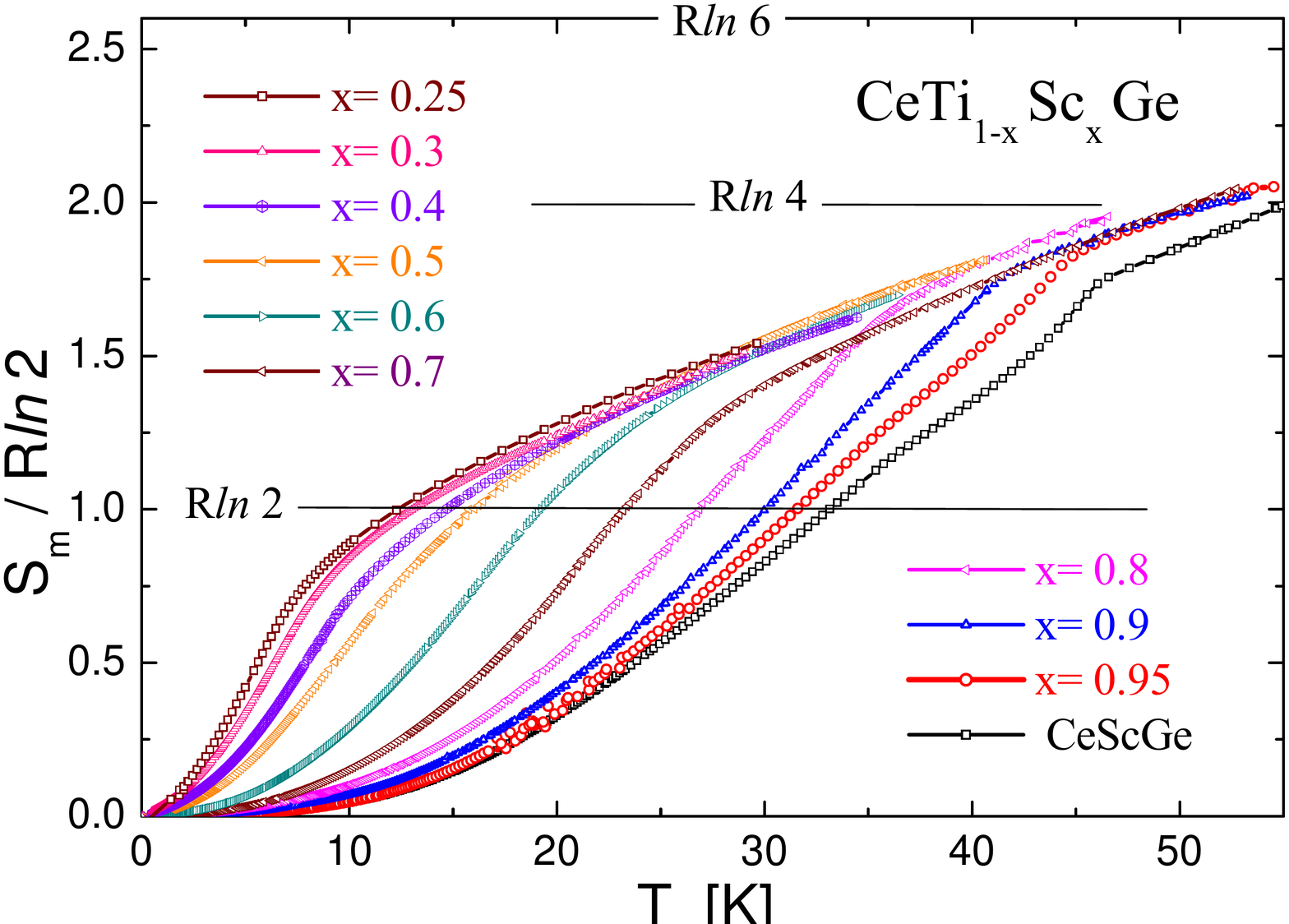}
\end{center}
\caption{(Color online) Thermal evolution of the magnetic
contribution to the entropy ($S_m$) normalized by $R\ln2$.}
\label{F9}
\end{figure}

The analysis of the thermal evolution of the magnetic contribution
to the entropy $S_m(T)$ is shown in Fig.~\ref{F9} normalized by
the value of a Kramer's doublet level: $R\ln 2$. This parameter
provides complementary information to better understand this
complex system. The alloy $x = 0.25$ with the lowest ordering
temperature ($T_C = 7$\,K) reaches $S_m= R\ln 2$ at $T = 11$\,K,
slightly above $T_C$, suggesting that only the GS doublet
contributes to the magnetic order. However, since in the
paramagnetic phase $S_m(T)$ increases continuously (i.e. without
showing any plateau around $R\ln 2$) one may infer that the low
energy tail of the first excited CEF level starts to contribute to
the entropy at quite low temperature. This fact confirms the CEF
level spectrum extracted in the previous subsection from the fit
of $C_m(T)$.

Although the contribution of the first excited doublet may be
marginal in the alloys with low Sc content, it becomes significant
for higher concentrations as $S_m (T_{ord})$ increases with
$T_{ord}(x)$. This feature becomes evident for $x \geq 0.5$ alloys
because $S_m (T_{ord})$ clearly exceeds $R\ln 2$. Notice that
around this concentration the AFM sets on. For $x=1$ it
practically reaches the value corresponding to two doublets, i.e.
$R\ln 4$, in agreement with the broad shoulder observed in
$C_m(T)/T$ and the value of $\Delta C_m(T_N)$ (see Fig.~\ref{F7}).

The comparison of the entropy distribution above and below
$T_{ord}$ (i.e. $S_m(T<T_C)$ and $S_m(T>T_C)$ respectively)
provides a hint to figure out the effective dimensionality of a
magnetic system. According to theoretic predictions
\cite{ProgPhys}, for two dimensional (2D) systems the entropy
accumulated between $T=0$ and $T=T_{ord}$ is similar to that
contained in the tail of $C_m(T>T_{ord})$. In this case, the
samples with $0.25\leq x \leq 0.5$ show $S_m(T<T_C)$ and
$S_m(T>T_C)$ values close to the prediction for a simple square
Ising lattice of spins 1/2. Such is not the case for samples
beyond the critical concentration (i.e. $x\geq 0.65$) for which
the shape of AFM transition tends to the characteristic $\Delta
C_m(T_N)$ jump of a 3D second order transition. Also this change
occurs around the modification of the magnetic character from FM
to AFM.

\subsection{Magnetic phase diagram}

\begin{figure}
\begin{center}
\includegraphics[angle=0, width=0.5 \textwidth, height=0.7\linewidth] {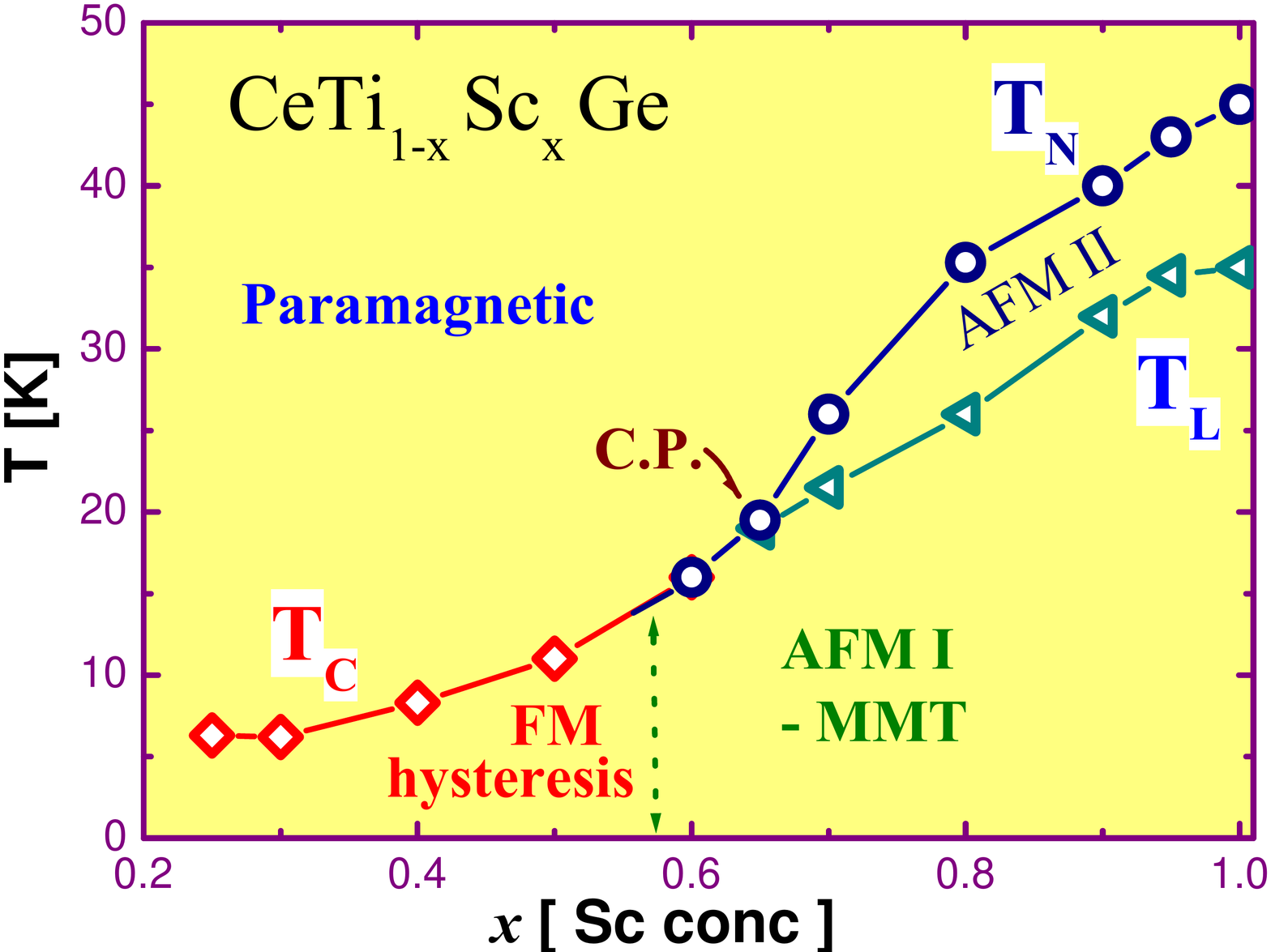}
\end{center}
\caption{(Color online) Sc concentration dependence of the
Magnetic phase diagram showing the transition temperatures
extracted from magnetic and thermal measurements (left axis)
within the range of the CeScSi-type structure. Saturation
magnetization $M_{sat}$ (right axis) of samples below the critical
point CP ($x=0.65$), compared with the effective magnetic moment
$M_{eff}$ of alloys beyond the CP, measured at $T=1.8$K and$\mu_0
H = 5$Tesla.} \label{F10}
\end{figure}

The magnetic characteristics of this system are summarized in the
phase diagram presented in Fig.~\ref{F10}. The two relevant
features are the FM to AFM change between $0.50\leq x \leq 0.60$
and the critical point CP at $x \approx 0.65$. There, an
intermediate phase sets in between $T_N(x)$ and $T_L(x)$. Along
the full concentration range different regions were identified: i)
between $0.25\leq x \leq 0.50$ a FM-GS, determined by $M$ vs $H$
hysteresis loops of Fig.~\ref{F6}a, show an increasing $M_{sat}$
that reaches its maximum value at $x=0.50$. ii) A continuous
change form FM to AFM occurs at $0.50 < x < 0.60$, evidenced by
the MMT with $H_{cr}$ arising from zero. $\rho(T)$ measurements
confirm this continuous change because the kink at $T=T_C$
progressively develops a characteristic AFM upturn (see
Fig.~\ref{F5}) due to the opening of a gap in a fraction of the
Fermi surface. iii) At the CP ($x \approx 0.65$) a weak transition
emerges at $T_L(x)$ below $T_N$ according to $C_m(T)$ and
$\chi(T)$ results. Notably, $T_N(x)$ rises continuously up to
$T_N=47$\,K (at $x = 1$) despite of the weakening of the Ce
effective magnetic moment.

All along the concentration range there is no indication for
increasing hybridization effects between $4f$ and conduction
electrons. On the contrary, some evidence for Kondo effect like
the increase in $\rho(T)$ towards low temperature and an
enhancement of $\gamma_0$ are observed on the Ti-rich side despite
the FM behavior. However, both parameters indicate the weakening
of this effect with concentration. Therefore, the high $T_N$
observed in CeScGe cannot be attributed to any hybridization
mechanism rather to an intrinsically strong RKKY interaction.

A general description of the magnetic evolution of this system can
be proposed by considering each Ce-double-layer as nearly 2D-FM
unit that weakly interacts with neighboring layers. The positive
and even growing $\theta_P(x)$ values support an increasing
$intra$-layer Ce-Ce neighbors FM interaction. Once the amount of
Sc-$3d^1$ electrons becomes relevant (at $x\approx 0.50$) the
$inter$-layers RKKY interaction takes over favoring an AFM
stacking of the double-layers within an anisotropic but 3D
scenario. The thermal population of the first CEF excited level
also contribute to the FM - AFM change because of the modification
of the effective moments.

Despite of $\approx 12\%$ difference between Ti and Sc Goldschmit
metallic radii ($r_{Ti}=1.46\AA$ and $r_{Sc}=1.64\AA$
respectively), the Ce-T spacing ($d_{Ce-T}$) only increases by
$\approx 2\%$ between CeTiGe ($d_{Ce-Ti}$ = 3.45\,$\AA$
\cite{Cheval}) and CeScGe ($d_{Ce-Sc}$ = 3.53\,$\AA$ \cite{Singh})
evaluated within the same CeScSi structure. This variation is in
agreement with the slight change ($\approx 1\%$) observed on the
'$c$' axis between $x=0.30$ and $x=1$. Within a {\it rigid sphere}
picture, this $d_{Ce-Sc}$ = 3.53\,$\AA$ and the sum of Sc and Ce
metallic radii ($r_{Sc}$ = 1.64\,$\AA$ and $r_{Ce}$ = 1.86\,$\AA$)
become comparable. Therefore, a significant increase of the
electronic overlap between Sc-$3d$ and Ce-$5d$ electrons can be
expected with the consequent strengthening of the
Ce-$inter$-layers RKKY interaction.

In this context one should notice that isotypic rare earth (R)
compounds of the RTGe family also present very high ordering
temperatures, like GdScGe (CeScSi type structure) that orders at
$T_C = 350$\,K \cite{G1}, more than a factor 7 higher than in
CeScGe. Similar values are observed for GdTiGe (of CeFeSi type
structure) with $T_N = 412$\,K \cite{G2}. These behaviors are in
clear contrast with itinerant cases like CeRh$_2$Si$_2$ and
CeRh$_3$B$_2$ \cite{Malik,Godart83}, for which the $T_N$ ratio
between GdRh$_2$Si$_2$ and CeRh$_2$Si$_2$ is only a factor $3$.

\section{Conclusions}

Apart from the significantly high ordering temperature of CeScGe
at $T_N = 47$\,K, the main magnetic characteristics shown by this
system are the continuous change from FM to AFM-I order and the
presence of a CP point (at $x=0.65$) associated to an intermediate
AFM-II phase. These modifications of the magnetic ground state
occur without affecting the local character of the Ce-$4f$
orbital.

The layer type of the crystalline structure seems to play a basic
role because it favors the formation of FM sheets involving two
neighboring Ce planes. At low Sc content ($0.25 \leq x \leq 0.50$)
the FM $intra$-plane RKKY interaction, resembling 2D-type order,
dominates the scenario. The continuous increase of the positive
$\theta_P$ temperature indicates that this Ce-Ce interaction even
enhances with concentration.

After reaching the maximum of $M_{sat}$ at $x=0.50$, the system
develops a MMT transition from $H_{cr} = 0$. Such a continuous
transition from a FM state to an AFM reveals the similar energy of
these competing phases, that can be described as a smooth
modification from a FM-stacking to an AFM-stacking of FM layers.
This change in the $inter$-layers RKKY interaction can be driven
by the variation of the electronic configuration of the
intermediaries atoms $Ti^{4+}$ and $Sc^{3+}$.

The increasing $inter$-layers interaction strengthens the 3D
character of the AFM phase evidenced by the $C_m(T)$ jump at the
magnetic transition. However, the intermediate phase AFM-II
suggests that the magnetic wave vector in the '$c$' direction does
not reach full commensuration at $T_N$. Nevertheless, the
discontinuity in the order parameter reflected in the $T_L$
transition decreases with growing the Sc content. The $S_m(T,x)$
evolution confirms the first excited CEF level increasing
contribution as $T_N$ becomes comparable to $\Delta_I$. Therefore,
the record high ordering temperature of CeScGe has to be
attributed to the convergence of different factors like: the Ce
double-layer structure of FM character, the increasing RKKY
interaction with Sc-$3d^1$ concentration and the vicinity of the
first CEF excited level. All these conditions are related with the
local characters of the Ce-$4f$ orbitals.

\end{document}